\documentclass[twocolumn]{aastex631}

\newcommand{\psr}{J0027$-$1956}
\newcommand{\psrdm}{20.869}
\newcommand{\psrdmerr}{0.005}
\newcommand{\psrnick}{J0036$-$1033}
\newcommand{\dmunit}{pc$\,$cm$^{-3}$}

\newcommand{\nullfraction}{77\%}
\newcommand{\ExB}{\ensuremath{{\bf E}\times{\bf B}}}

\usepackage{amsmath}

\shorttitle{Pulsar with unusual sub-pulse drifting}
\shortauthors{McSweeney et al.}
\graphicspath{{./}{figures/}}

\begin{document}

\title{Independent discovery of a nulling pulsar with unusual sub-pulse
drifting properties with the Murchison Widefield Array}

\author[0000-0001-6114-7469]{Samuel J. McSweeney}
\affiliation{International Centre for Radio Astronomy Research, Curtin University, Bentley, WA 6102, Australia}

\author[0000-0002-8383-5059]{N. D. Ramesh Bhat}
\affiliation{International Centre for Radio Astronomy Research, Curtin University, Bentley, WA 6102, Australia}

\author[0000-0001-8982-1187]{Nicholas A. Swainston}
\affiliation{International Centre for Radio Astronomy Research, Curtin University, Bentley, WA 6102, Australia}

\author[0000-0001-7801-9105]{Keegan R. Smith}
\affiliation{International Centre for Radio Astronomy Research, Curtin University, Bentley, WA 6102, Australia}

\author[0000-0002-6631-1077]{Sanjay Kudale}
\affiliation{National Centre for Radio Astrophysics, Tata Institute of Fundamental Research, Pune 411 007, India}

\author[0000-0002-4203-2946]{Paul Hancock}
\affiliation{Curtin Institute for Computation, Curtin University, GPO Box U1987, Perth, 6845, WA, Australia}

\author[0000-0003-2519-7375]{Willem van Straten}
\affiliation{Institute for Radio Astronomy \& Space Research, Auckland University of Technology, Private Bag 92006, Auckland 1142, New Zealand}

\author[0000-0002-9618-2499]{Shi Dai}
\affiliation{Western Sydney University, Locked Bag 1797, Penrith South DC, NSW 1797, Australia}

\author[0000-0002-7285-6348]{Ryan M. Shannon}
\affiliation{Centre for Astrophysics and Supercomputing, Swinburne University of Technology, Hawthorn, VIC, 3122, Australia}
\affiliation{ARC Centre of Excellence for Gravitational Wave Discovery (OzGrav)}

\author[0000-0002-8195-7562]{Steven J. Tingay}
\affiliation{International Centre for Radio Astronomy Research, Curtin University, Bentley, WA 6102, Australia}

\author[0000-0003-2756-8301]{Melanie Johnston-Hollitt}
\affiliation{Curtin Institute for Computation, Curtin University, GPO Box U1987, Perth, 6845, WA, Australia}

\author[0000-0001-6295-2881]{David~L.~Kaplan}
\affiliation{Center for Gravitation, Cosmology, and Astrophysics, Department of Physics, University of Wisconsin-Milwaukee, P.O. Box 413, Milwaukee, WI 53201, USA}

\author[0000-0003-4077-6244]{Mia Walker}
\affiliation{International Centre for Radio Astronomy Research, Curtin University, Bentley, WA 6102, Australia}



\begin{abstract}

We report the independent discovery of PSR \psr{} with the Murchison Widefield Array (MWA) in the ongoing Southern-sky MWA Rapid Two-meter (SMART) pulsar survey.
\psr{} has a period of $\sim1.306$ s, a dispersion measure (DM) of $\sim\psrdm\,$\dmunit, and a nulling fraction of $\sim$~\nullfraction{}.
This pulsar highlights the advantages of the survey's long dwell times ($\sim80\,$min), which, when fully searched, will be sensitive to the expected population of similarly bright, intermittent pulsars with long nulls.
A single-pulse analysis in the MWA's 140-170 MHz band also reveals a complex sub-pulse drifting behavior, including both rapid changes of the drift rate characteristic of mode switching pulsars, as well as a slow, consistent evolution of the drift rate within modes. In some longer drift sequences, interruptions in the otherwise smooth drift rate evolution occur preferentially at a particular phase, typically lasting a few pulses.
These properties make this pulsar an ideal test bed for prevailing models of drifting behavior such as the carousel model.

\end{abstract}

\section{Introduction} \label{sec:intro}

Pulsars are rapidly rotating neutron stars, the sites of some of the highest energy physical processes in the universe, owing to the very strong gravitational and magnetic fields that surround them.
Discovered first via their lighthouse-like beam of radio emission \citep{Hewish1968}, which we detect as a series of regularly spaced pulses, it was quickly realised that the physics governing the relativistic plasma that generates this beam was not well understood---a state of affairs that persists to the present day \citep[e.g.][]{Melrose2021}.
The uniqueness of each pulsar's emission signature provides a wealth of information that can be used to test proposed emission mechanisms, and it has been proven again and again over the past several decades that in-depth studies of individual pulsars can provide valuable clues for understanding the population as a whole.
This, along with tests of general relativity \citep[e.g.][]{Kramer2006,Miao2021}, pulsar timing arrays for gravitational wave detection \citep[e.g.][]{Manchester2013}, pulsar braking and magnetospheric dynamics \citep[e.g.][]{Gao2017,Wang2020}, the neutron star equation of state \citep[e.g.][]{Demorest2010,Antoniadis2013,Pang2021}, and other pulsar science applications, motivates efforts to find new pulsars via large scale pulsar surveys, which continue to be conducted up to the present day.

One of the most recent such surveys is currently under way at the Murchison Widefield Array (MWA), a latest-generation aperture array telescope located in Western Australia \citep{Tingay2013}.
With its Voltage Capture System \citep[VCS;][]{Tremblay2015} and offline tied-array beamforming software \citep{Ord2019,McSweeney2020}, it has proven to be a powerful instrument for investigating a variety of pulsar phenomena such as single-pulse studies \citep{McSweeney2017}, spectral analyses \citep{Meyers2017}, studies of interstellar medium (ISM) propagation effects, and related timing applications \citep{Bhat2018,Kaur2019}.
With the advent of the Phase 2 upgrade of the MWA \citep{Wayth2018}, which allows a compact configuration of 128 tiles (i.e. antenna elements) with short baselines ($\lesssim300\,$m), a pulsar survey became computationally feasible, owing to the relatively large angular size of the tied-array beam ($\sim23^\prime$).
The Southern-Sky MWA Rapid Two-metre (SMART) pulsar survey (Bhat et al. \textit{in prep}) was conceived and data collection began in late 2018.
At present, $\sim70\%$ of the data have been collected, with the initial, first-pass (``shallow'') processing being limited to the first ten minutes (out of the full 80 minutes) of each observation.
This strategy is beginning to pay off; the discovery of PSR \psrnick{}, a low-luminosity, high Galactic latitude pulsar ($b \approx -73^{\circ}$), was reported in \citet{Swainston2021}.

In this paper, we report the MWA's independent discovery of PSR \psr{} in the shallow pass of the SMART survey.
The pulsar was originally detected in 2018 in the Green Bank Northern Celestial Cap (GBNCC) pulsar survey \citep{Stovall2014}.
Unlike \psrnick{}, \psr{} is sufficiently bright to be detected in single pulses, and was found to have a large nulling fraction.
Closer inspection of \psr{}'s single pulses revealed that it belongs to the class of sub-pulse drifters, a phenomenon first noted in the earliest days of pulsar research \citep{Drake1968}, and which has historically been strongly linked to magnetospheric phenomena, and thence, to the as-yet poorly understood radio emission mechanism \citep{Ruderman1975,Rankin1986,Deshpande1999,McSweeney2019a}.
``Sub-pulse'' signifies discrete bursts of emission, usually much narrower than the average pulse profile, and ``drifting'' refers to the systematic way that sub-pulses arrive earlier or later in time in successive rotations.
Sub-pulse drifting is easily detected when viewing the pulses in a stack (pulse number vs rotation phase), in which associated sets of sub-pulses form diagonal \textit{drift bands} that stretch across the on-pulse window.
The \textit{drift rate} is defined as the reciprocal of the slope of the drift bands, often expressed in units of degrees (of rotation) per pulse period ($^\circ/P$).

Many sub-pulse drifters, e.g., B0809$+$74 and B0943$+$10, are known to have very stable drift rates \citep[e.g.][]{Taylor1971,Deshpande1999}.
This stability partly motivated the original carousel model, put forward by \citet{Ruderman1975}, which associates the sub-pulses with spatially discrete bursts of electrical discharges (``sparks'') in regions of charge depletion just above the stellar surface near the magnetic poles.
According to this model, the sparks are driven by \ExB{} drift, with the observed stability of the drift rate being directly inherited from the theoretical stability of the electric and magnetic fields at the spark locations.
Therefore, pulsars that exhibit anything more complicated than a single, stable drift rate deserve special attention, since they require modifications of, or at least extensions to, the basic carousel model.
Such pulsars exist, and several extensions have been proposed over the years to account for them.
For example, \citet{Gil2000} suggest that a quasi-central spark can account for (non-drifting) core components in profiles, and the well-known phenomenon of bi-drifting may be explained by presence of an inner annular gap \citep{Qiao2004}, an inner acceleration gap \citep{Basu2020}, or non-circular spark motions \citep{Wright2017}.
Such extensions are typically developed to explain specific drifting behaviors that are observed in a relatively small subset of pulsars, and there still lacks a single, comprehensive theory that can describe all drifting behaviors.

Pulsars that both show complicated drifting behavior and are bright enough for single pulse analysis, are relatively rare.
Even a cursory glance at the drift bands of \psr{} show that its drifting behavior is indeed complex and bright, making it a welcome addition to this category of pulsar.
\psr{} is also found to contain null sequences, a phenomenon which is known to be connected to sub-pulse drifting, originally noticed by \citet{Lyne1983}, and further attested by recent studies of PSRs J1727$-$2739 \citep{Wen2016}, B1819$-$22 \citep{Janagal2022}, and the Vela pulsar \citep{Wen2020}.

This paper presents an analysis of the nulling and drifting behavior of \psr{} observed with the MWA.
Section \S\ref{sec:observations} describes the MWA observation in which the pulsar was originally found and the small number of archival MWA observations in which it was subsequently detected during follow up of the original candidate.
Analysis of the pulsar's nulling and drifting properties follows in Section \S\ref{sec:analysis}, which are discussed and compared to other pulsars in Section \S\ref{sec:discussion}.
We conclude with a short summary in Section \S\ref{sec:conclusions}.

\section{Observations}
\label{sec:observations}


PSR \psr{} was discovered in one of the observations made as part of the SMART survey (Obs ID 1226062160).
A full description of the survey parameters and observational set up will be presented in an upcoming publication \citetext{Bhat et al. in prep}.
Many details are also described in the paper reporting the first SMART pulsar discovery \citep{Swainston2021}, for which the setup was identical, but the essential details are summarised here.
The VCS delivers Nyquist-sampled dual-polarisation voltages for 128 MWA tiles ($4\times4$ cross-dipoles) at a rate of $100\,\mu$s and an instantaneous bandwidth of $30.72\,$MHz, consisting of $3072\times10\,$kHz individual channels.
Apart from the discovery observation, the other five observations presented in this work were taken as part of the follow-up timing campaign for \psrnick{} \citep{Swainston2021}.
All six observations (summarised in Table \ref{tab:observations}) were made in the frequency range $138.88$-$169.6\,$MHz.

\begin{deluxetable}{cccccc}
\tabletypesize{\scriptsize}
\tablewidth{0pt}
    \tablecaption{MWA Observations of PSR \psr{} \label{tab:observations}}
\tablehead{
    \colhead{MJD} & \colhead{Obs ID} & Length & \multicolumn{2}{c}{Number of pulses} \\
    & & (mins) & Null & Not null
}
\startdata
    58434 & 1226062160 & 82 & 2971 & 788 \\
    58991 & 1274143152 & 20 & 556  & 358 \\
    59002 & 1275094456 & 20 & 907  &  10 \\
    59002 & 1275172216 & 20 & 901  &  16 \\
    59003 & 1275178816 & 20 & 633  & 281 \\
    59094 & 1283104232 & 30 & 829  & 545 \\
    Total &           & 192 & 6797 & 1998
\enddata
\tablecomments{All observations are in the $140$-$170\,$MHz band, with $100\,\mu$s/$10\,$kHz resolutions.}
\end{deluxetable}

\begin{figure}[!th]
    \centering
    \includegraphics[width=\linewidth]{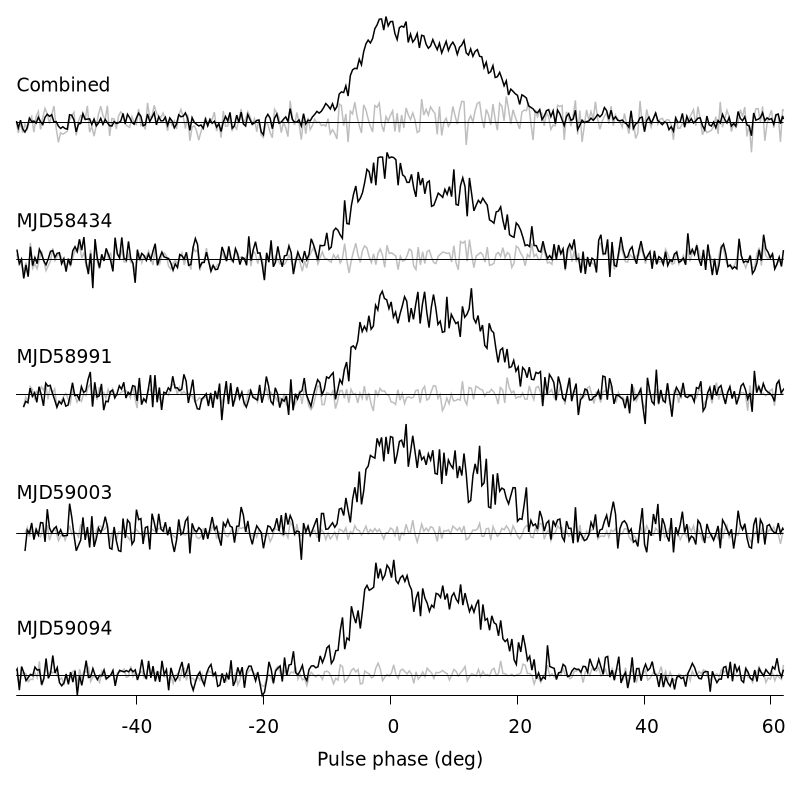}
    \caption{Peak-normalised mean profiles (black) of four of the observations, and the total mean profile (top panel) including all four observations, with a time resolution of $\sim1.3\,$ms. The two observations on MJD 59002 are not included here as they consist almost entirely of noise. The profiles show a barely resolved double peak structure. The gray lines show the profiles made up only of pulses classified as nulls.}
    \label{fig:profiles}
\end{figure}

Each observation was individually calibrated using observations of 3C444 taken within $\sim2$ hours of the respective target observations.
The calibration solutions were obtained using the Real Time System software \citep{Mitchell2008}.
After calibration, the voltages were used to form a tied-array beam in the direction of the pulsar (the localization procedure is discussed in \S\ref{sec:localization}), using the bespoke beamforming software described in \citet{Ord2019}.
The MWA is located in a very radio quiet location, and no significant radio frequency interference was found in the beamformed data.
After determining the best dispersion measure (DM) and period (see \S\ref{sec:period}), de-dispersion and folding were performed with {\tt DSPSR} \citep{VanStraten2011b} and {\tt PSRCHIVE} \citep{Hotan2004} to form single pulse archives, from which Stokes I pulse stacks were created.
Profiles of each detection, as well as an integrated profile incorporating the data from all four observations, is shown in Fig \ref{fig:profiles}.
Subsequent analysis of the pulse stacks is described in \S\ref{sec:analysis}.

Although this pulsar was detected in the shallow pass of SMART, it was subsequently noted to have been reported in 2018 as a candidate in the GBNCC pulsar survey \citep{Stovall2014}.
Later re-processing of data from the 70-cm Parkes Southern Pulsar Survey  \citep{Manchester1996} also revealed a weak detection that was not previously reported.

Follow-up of the initial MWA detection with the full 80 minutes, as well as in other archival MWA observations, revealed that the pulsar has a very large nulling fraction ($\sim$~\nullfraction{})\footnote{The nulling fraction estimate comes only from analysis of the MWA observations.}, with some nulling sequences (excluding occasional intermittent pulses) potentially exceeding $20\,$min (i.e. the length of the first observation on MJD 59002, which contained \textit{only} a few intermittent pulses, as shown in Fig. \ref{fig:maxima}).
A full analysis of the nulling properties is presented in Section \S\ref{sec:nulling} below.
Despite the fact that \psr{} was serendipitously discovered in the first ten minutes, it may well be the harbinger for a population of relatively bright, intermittent pulsars which the SMART survey, with its $80$-minute dwell times, is well-suited to detect.
A similar strategy is employed in the Low Frequency Array (LOFAR) Tied-Array All-Sky Survey \citep[LOTAAS;][]{Sanidas2019} and the low latitude ($|b| \le 5^\circ$) segment of the High Time Resolution Universe survey \citep[HTRU;][]{Keith2010}, which use dwell times of approximately one hour.

\subsection{Localization}
\label{sec:localization}

The discovery observation (Obs ID 1226062160) was taken when the MWA was in its \textit{compact} configuration \citep[see][]{Wayth2018}, giving an effective tied-array beam size of $\sim23^\prime$ at the central frequency of $155\,$MHz.
As per the SMART survey design, these beams are tiled across the field of view, each of which is searched for periodic candidates.
The compact configuration includes many redundant baselines, and the tied-array beam for the compact configuration is complex.
Consequently, the pulsar was detected in multiple adjacent beams, including a boresight beam that yielded the highest signal-to-noise, and several nearby grating lobes.
The ability to exploit grating lobe detections for fast-tracking candidate confirmation and localization will be discussed in the survey description paper \citetext{Bhat et al. in prep}.

The other five observations were taken when the MWA was in its extended configuration, giving a tied-array beam size of $\sim3^\prime$ at $155\,$MHz.
Localization within the four observations in which the pulsar was bright enough followed the identical procedure as used in \citet{Swainston2021}, i.e.\ gridding the candidate positions with overlapping tied-array beams and using signal-to-noise and the estimated beam shape to produce a likelihood map of the source position.
It is known that the ionosphere is capable of shifting the apparent position of sources by an appreciable fraction ($\sim10$-$30\%$) of the tied-array beam, and that these offsets are not necessarily corrected by the calibration process when the calibration observations are separated from the target observation, both spatially and temporally \citep{Swainston2022}.
Nevertheless, the localization afforded by the extended array observations was sufficiently precise (a few arcminutes) to warrant follow-up with a high-resolution imaging telescope.
Accordingly, images made with Band 3 ($300$-$500\,$MHz) data from the upgraded Giant Metrewave Radio Telescope \citep[uGMRT;][]{Gupta2017}, obtained under Director's Discretionary Time (DDTC185), enabled us to confirm the pulsar's position to an accuracy of $\sim1^\prime$.
The right ascension and declination of the final best position, consistent with both the uGMRT imaging and the MWA extended configuration observations, are 00h26m36s and $-$19$^\circ$55'59''.

\subsection{Determination of Period and Dispersion Measure}
\label{sec:period}

The limited number of available detections make obtaining a timing solution impossible at this early stage.
Further follow up observations with the uGMRT and the Parkes 64-m radio telescope (also known as \emph{Murriyang}), as well as continued processing of archival MWA observations, are currently underway. These observations, along with the full timing solution that they are expected to yield, will be reported in a future publication.

Lacking a timing solution, we used PSRCHIVE's \texttt{pdmp} routine to determine the best period ($P$) and DM for each observation independently. This routine performs a brute-force grid search in both period and DM parameter space, and returns the period and DM that yields the highest profile signal-to-noise. The weighted averages of both period and DM (respectively) over the four observations were calculated, with the uncertainties reported by \texttt{pdmp} being used to weight the individual measurements.
The values thus derived were $P = 1.306150 \pm 0.000005\,$s and $\text{DM} = \psrdm \pm \psrdmerr\,$\dmunit.
This simple estimate does not take the period derivative into account (in fact, it implicitly assumes $\dot{P} = 0$), and we note that a sufficiently large $\dot{P}$ would cause the period to drift significantly over the time period spanned by our four MWA observations ($\sim1.5\,$yr).
However, the pulse stacks (e.g.\ Fig. \ref{fig:pulsestacks}) folded on the above period showed no detectable slope, which we estimate would be discernible if the pulse window changed by more than $20\,$ms over the course of the observation.
Given this, we place a conservative upper limit of $|\dot{P}| \le 10^{-13}$\,s\,s$^{-1}$.

Any error in the folding period also naturally translates into a systematic error in the sub-pulse drifting analysis presented below, e.g.\ the drift rate.
However, the fractional uncertainty of the period is much smaller than those of the sub-pulse drifting analysis, which are dominated by the stochastic and bursty nature of the sub-pulses themselves.
Hence, in the following analysis, we neglect this systematic error, which is not expected to affect the main results.

\section{Analysis and Results}
\label{sec:analysis}

\subsection{Nulling properties}
\label{sec:nulling}

\begin{figure*}[t]
    \includegraphics[width=\textwidth]{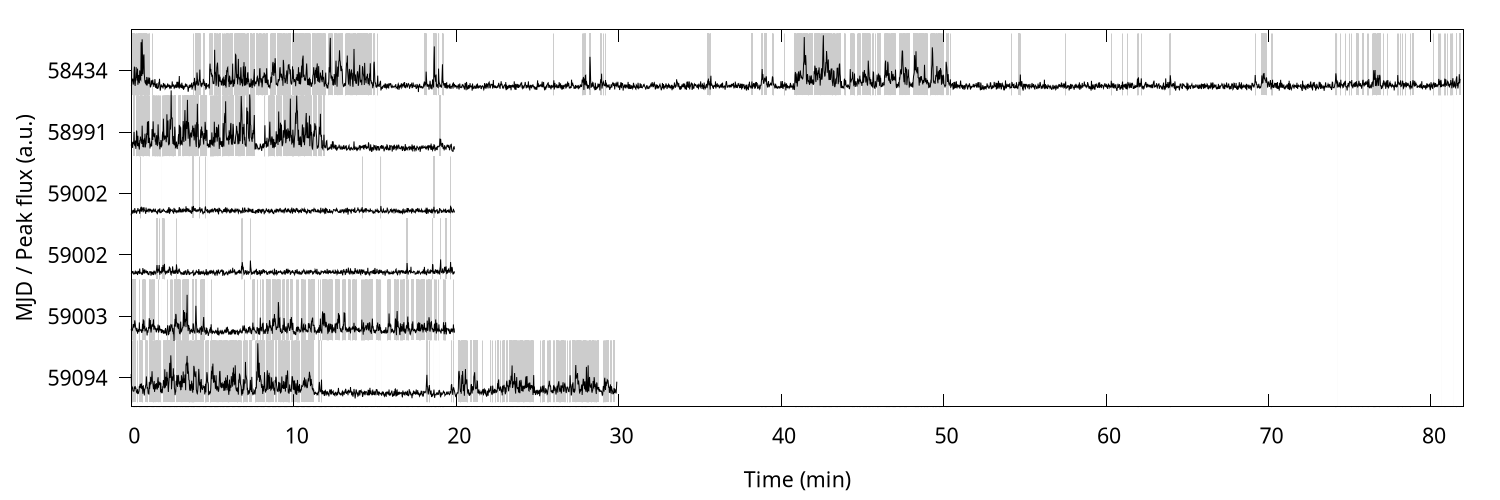}
    \caption{Peak flux density within the pulse window as a function of time for six MWA observations (MJDs given along the $y$-axis). The shaded gray regions show detected subpulses that are included in the nulling fraction estimation, as explained in the text. The peak fluxes were measured after smoothing the time series with a Gaussian filter ($\sim2\,$ms wide, the approximate width of a sub-pulse) to suppress the noise contribution and accentuate the contrast between nulling and burst sequences. Note that the low-level peaks in the $\sim70$-$80\,$mins time range of the MJD 58434 and in the second MJD 59002 observation are single pulses faintly visible in the pulse stack, but which are faint because the pulsar was moving out of the MWA's primary beam.}
    \label{fig:maxima}
\end{figure*}

The emission from \psr{} appears either as long ($\gtrsim 10\,$min) burst sequences exhibiting phase-modulated sub-pulse drifting, or as short ($\lesssim 20\,$s) bursts interspersed throughout otherwise long (up to $25\,$min) nulling periods.
Both types of emission are visible in Fig. \ref{fig:maxima}, in which the peak flux density within the pulse window of each pulse is plotted as a function of time for all four MWA observations.

To estimate the nulling fraction, we first identified which pulses contained sub-pulses within the pulse window (this is a semi-automated procedure described in \S\ref{sec:drifting_analysis}).
Any pulse without a sub-pulse is then counted as a null, even if it is nested within a burst sequence.
Under this definition, we report a nulling fraction of $\sim$~\nullfraction{},
calculated using all six MWA observations (refer to Table \ref{tab:observations}).
Note that if the two observations taken on MJD 59002 are excluded, the nulling fraction would be $\sim$~72\%.

\begin{figure*}[p]
    \centering
    \includegraphics[width=0.48\textwidth]{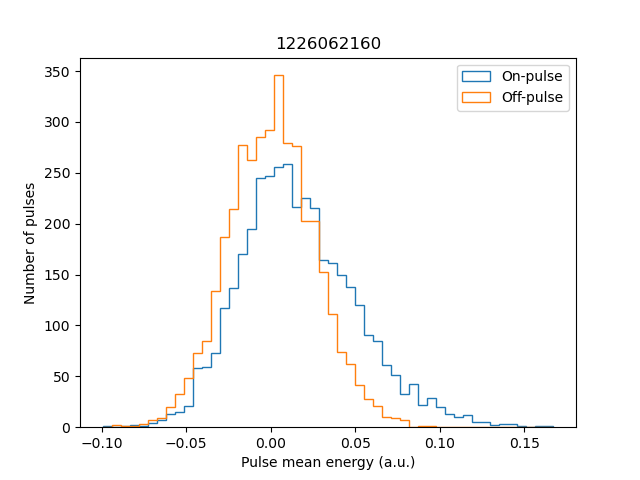}
    \includegraphics[width=0.48\textwidth]{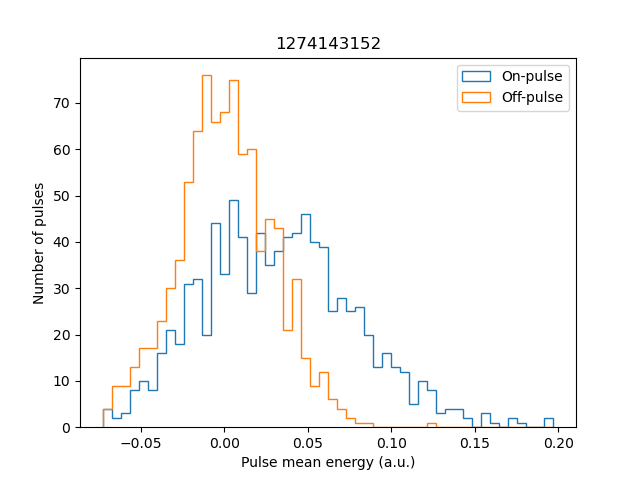} \\
    \includegraphics[width=0.48\textwidth]{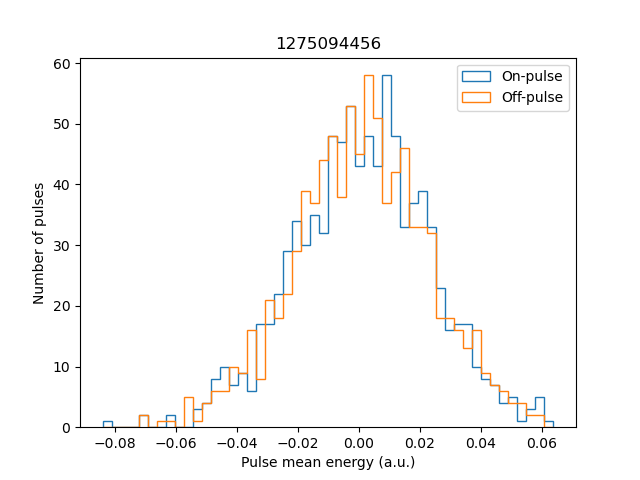}
    \includegraphics[width=0.48\textwidth]{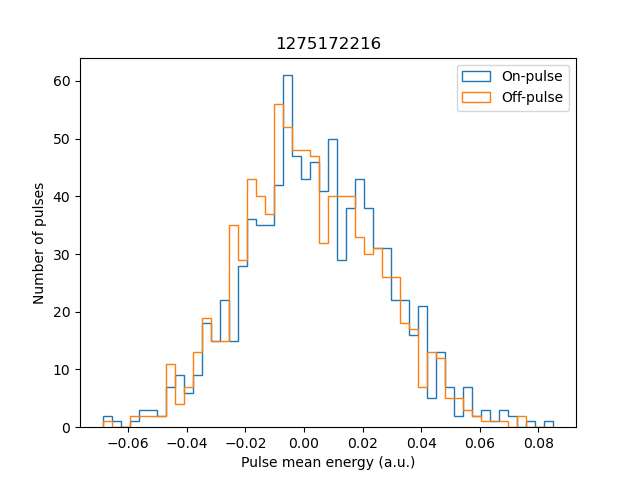} \\
    \includegraphics[width=0.48\textwidth]{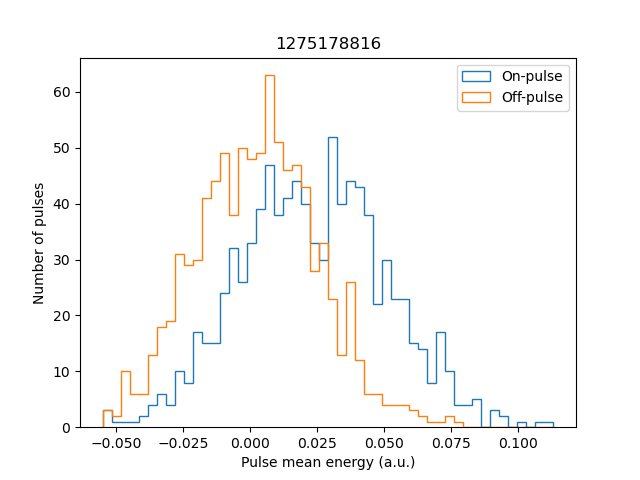}
    \includegraphics[width=0.48\textwidth]{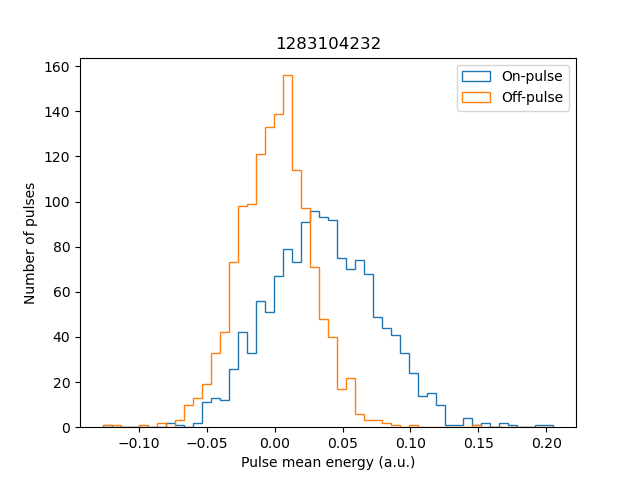}
    \caption{The pulse energy histograms for each of the six observations, for matched-size on- and off-pulse regions of the pulse stacks. The ObsIDs are given above each plot. The on-pulse histograms include all pulses, regardless of whether they were classified as nulls by the method described in the text.}
    \label{fig:nullhist}
\end{figure*}

This working definition of a null may misclassify pulses with broad emission spread out over the pulse window, even if its integrated flux density is statistically significant.
For comparison, we present the pulse energy histograms of both on- and off-pulse regions of the pulse stacks for each of the six observations in Fig. \ref{fig:nullhist}.
The similarity between the on- and off pulse distributions for observations 1275094456 and 1275172216 indicate that any broad emission that is present falls well below the MWA detection threshold.
Despite this, the uncertainty in our estimate of the nulling fraction will still be dominated by the lack of a statistically significant number of burst sequences.

The error on this estimate, however, will be dominated by the small number of burst/null sequences captured in these six observations.
Therefore, the true nulling fraction may ultimately prove to be much lower or higher than the $\sim$~\nullfraction{} quoted here.
Ongoing follow-up observations (and a more complete search through archival MWA data) will allow us to place much stronger constraints on the nulling fraction, and will be reported in a future publication.

\subsection{Drifting behavior}
\label{sec:drifting_analysis}

\begin{figure*}[!t]
    \centering
    \includegraphics[width=0.2955\textwidth]{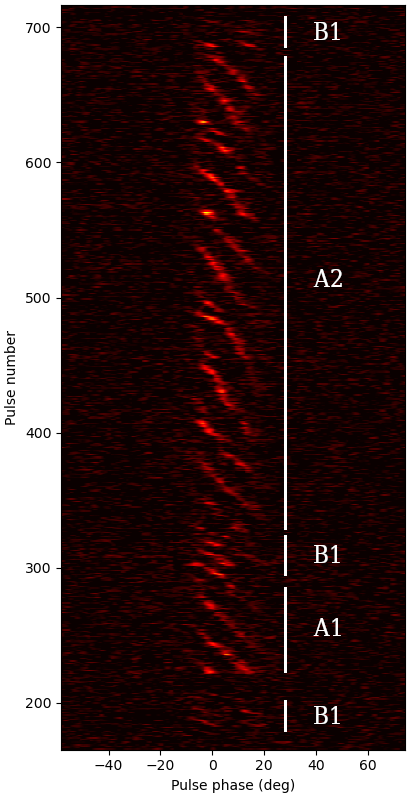}
    \includegraphics[width=0.3\textwidth]{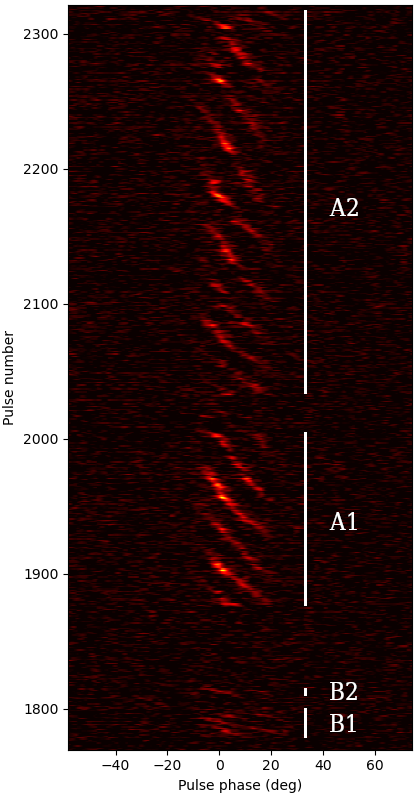}
    \includegraphics[width=0.302\textwidth]{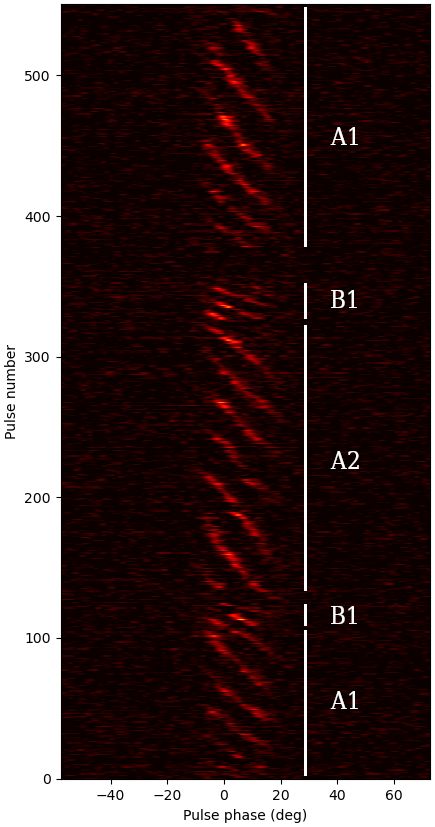}
    \caption{Three sets of 551 consecutive pulses from observations on MJD 58434 (left, middle) and 58991 (right), with labels identifying the classification of drift modes. To increase the visual contrast, the pulses have been smoothed with a $\sigma = 1^\circ$ Gaussian and the colour ranges manually adjusted (the flux densities have arbitrary units, and are not shown here). The origin of the phase axis has been set for each observation independently to correspond approximately to the peak of the average profile. See the main text for a description of the modes.}
    \label{fig:pulsestacks}
\end{figure*}

During burst sequences, the single pulses form distinct phase-modulated drift bands that are always oriented such that sub-pulses associated with the same drift band appear at earlier rotation phases over time \citep[`positive drifting' in the classification scheme of][]{Basu2019a}.
The drift rate, however, is highly variable, as can be seen in the example pulse stacks in Fig \ref{fig:pulsestacks}.
Sometimes the drift rate appears to change abruptly, reminiscent of the drift mode changes observed in several other pulsars \citep[e.g.][]{Kloumann2010,Rankin2013,Joshi2017,McSweeney2019a}.
However, apart from these abrupt changes, the drift rate is seen to evolve gradually, as does the ``vertical spacing'' between them, denoted by $P_3$ (see, e.g., the sequences marked ``A1'' in Fig \ref{fig:pulsestacks}).
The most dramatic examples of this slow evolution show a difference of more than a factor of two between the value of $P_3$ at the beginning and end of a burst sequence.

The manner in which the drift rate and $P_3$ evolve make the clean division of burst sequences into distinct drift modes difficult.
Nevertheless, the most economical description of the observed drifting behavior is that it consists primarily of two modes, which we label A and B.
We further divide these into two subcategories (A1/A2 and B1/B2) which depend on qualitative properties of their appearance and context.
This categorisation scheme is described below.

\subsubsection{Mode classification}

Mode A is characterised by slowly evolving $P_3$ values that can range from $\sim20$ to $\sim55\,P$, and drift rates in the range $-0.3$ to $-0.7^\circ/P$.
Most commonly, $P_3$ is seen to \textit{increase} over time, as the examples in Fig. \ref{fig:pulsestacks} show.
However, there is at least one clear instance in the (MJD 59094 observation where $P_3$ decreases from approximately $35$ to $24\,P$ over the course of 170 rotations.

Mode A sequences can last anywhere from a few tens to a few hundreds of pulses.
For longer sequences, Mode A is easily identifiable from the relatively large $P_3$, but for shorter sequences, only one or two drift bands are visible, and therefore $P_3$ cannot be measured directly.
In these cases, a positive identification of Mode A is based on the drift rate, which can be measured even for single drift bands.

The distinction between the subcategories A1 and A2 is based primarily on the level of organisation of the drifting behavior.
Sequences whose drift bands follow smooth (albeit curved) tracks with little deviation are classified as A1.
On the other hand, A2 sequences show frequent interruptions, both in terms of short-duration nulls as well as rapid, but temporary, deviations in the drift rate.
These interruptions are observed to last between $\sim$~5-10 pulses, but afterwards the drift bands reappear at the rotation phase that one would expect if the interruptions had not occurred.
These are further discussed in \S\ref{sec:P2P3DR}.

Mode B sequences are relatively short (up to 30 pulsar rotations in duration), and have small $P_3$ values ($\sim10\,P$) and large drift rates ($\sim-1.2^\circ/P$).
Longer sequences ($\gtrsim 10$ rotations) are found to occur most commonly nearby to Mode A sequences; these are classified as B1.
Shorter sequences ($\lesssim 10$ rotations) may be found anywhere, even in the midst of long, otherwise uninterrupted null sequences.
As with Mode A sequences, those sequences which are too short in duration to admit more than one or two drift bands may still be positively identified by their drift rate.

\subsubsection{Drift band modelling}

The pulse stacks in Fig \ref{fig:pulsestacks} clearly show that the slow evolution of the drift rate is nevertheless ``fast'' enough that even individual drift bands (which can span in excess of 50 pulses) can show significant curvature.
Following the lead of \citet{Lyne1983}, we therefore modelled the drift bands with an empirical function that assumes an exponential decay rate for the drift rate, $D$, following a null:
\begin{equation}
    D = D_0 e^{-p/\tau_r} + D_f,
    \label{eqn:exp_dr}
\end{equation}
where $D_0$ is the difference between the asymptotic drift rate, $D_f$, and the drift rate at the onset of the drift sequence; $p$ is the number of pulses since that onset; and $\tau_r$ is the drift rate relaxation time (in units of the rotation period).

The fitting procedure is carried out as follows.
The beginning and ending pulses of a drift sequence were manually identified by visual inspection of the pulse stack.
Sub-pulses were then identified by first smoothing the pulses with a Gaussian kernel of width $\sim3.6\,$ms (i.e.\ $1^\circ$ of pulsar rotation, the approximate width of a sub-pulse), and identifying peaks above a certain flux density threshold.
The threshold is chosen to be the minimum pixel value for which no sub-pulses are identified in the off-pulse region throughout the entire pulse stack.
Each sub-pulse in the drift sequence is then assigned a drift band number, and fitted to the following functional form of the sub-pulse phases (using SciPy's \texttt{curve\_fit} method).
This function is obtained by considering the drift rate as the rate of change of sub-pulse phase with pulse number, and integrating Eq.~\eqref{eqn:exp_dr}:
\begin{equation}
    \varphi = \tau_r D_0 \left(1 - e^{-p/\tau_r}\right) + D_f p + (\varphi_0 + P_2 d),
    \label{eqn:exp_phase}
\end{equation}
where $\varphi$ is the phase of a sub-pulse; $\varphi_0$ is an initial reference phase; $d$ is the (integer) drift band number; and $P_2$ is the longitudinal spacing between successive drift bands.
In total, this model has five free parameters, $D_0$, $D_f$, $\tau_r$, $\varphi_0$, and $P_2$, of which the expression $\varphi_0 + P_2 d$ defines the pulse phase at $p = 0$.

During the fitting, no restriction is placed on the sign of $\tau_r$.
A positive value would mean that the drift rate eventually asymptotes to $D_f$, whereas a negative value would mean that the drift rate is increasing exponentially from an initial drift rate (i.e.\  $D\rightarrow D_f$ as $p\rightarrow -\infty$).
Unless the relaxation time is much less than the duration of the drift sequence, $\tau_r$ is highly covariant with the other parameters, and the model is not capable of distinguishing between the two scenarios with any confidence.
Given this, the primary utility of the empirical model is its ability to predict other measurable quantities such as $P_2$, $P_3$, and the drift rate, which are less sensitive to the precise functional form used.

\subsubsection{$P_2$, $P_3$, and drift rate}
\label{sec:P2P3DR}

The drift band fitting model defined in Eqs. \eqref{eqn:exp_dr} and \eqref{eqn:exp_phase} can be used to extract information about $P_2$, $P_3$, and the drift rate, $D$.
$P_2$, which is assumed to be constant everywhere throughout a drift sequence, is fit independently for each drift sequence.
$P_3$ and the drift rate are defined continuously over all pulse numbers (even fractional ones) by virtue of the fact that the drift bands are modelled as continuous functions on the pulse stack, although we caution that these quantities are only actually measured within the pulse window, at intervals of a near-whole number multiple of the rotation period.
Eq.~\eqref{eqn:exp_dr} gives the drift rate, and one can then derive $P_3 = P_2/|D|$ directly.
The assumption of constant $P_2$ (for a given drift sequence) is justified by virtue of the goodness-of-fit of the models to the drift bands.

\begin{figure*}[!tp]
    \centering
    \includegraphics[width=\textwidth]{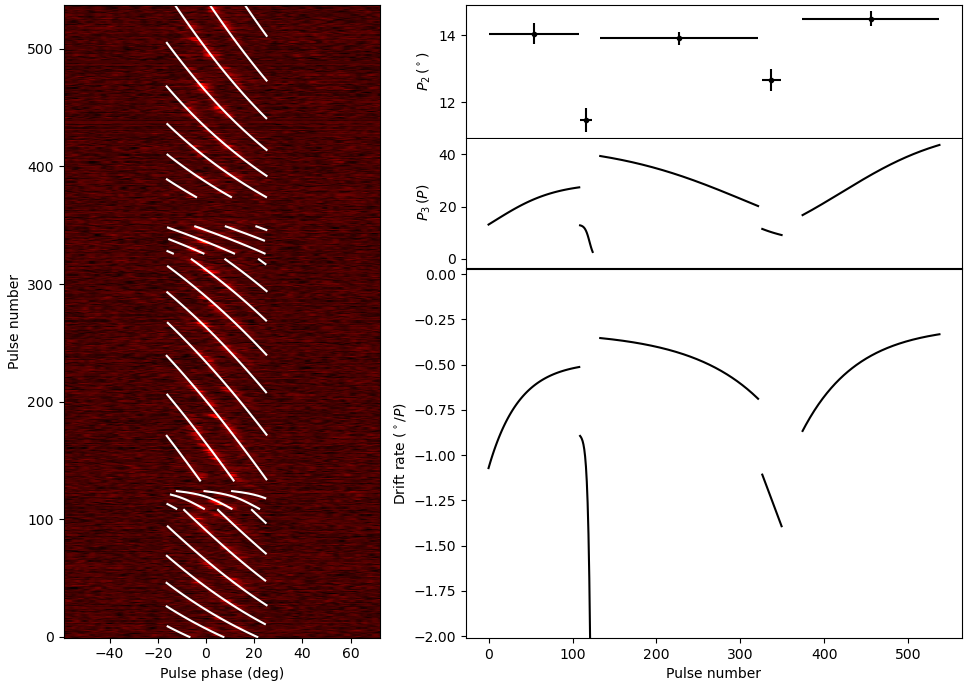}
    \caption{\textit{Left:} The pulse stack of MJD 58991 (right panel of Fig \ref{fig:pulsestacks}), with the drift bands modelled with an exponentially decaying drift rate. \textit{Right:} The $P_2$, instantaneous $P_3$, and instantaneous drift rates derived from the model. In the $P_2$ panel, the vertical error bars indicate the error on the fitted $P_2$ value, and the horizontal error bars only indicate the extent of the associated drift sequence.}
    \label{fig:with_model}
\end{figure*}

Fig. \ref{fig:with_model} shows the result of the drift band fitting to the MJD 58991 observation, alongside the values of $P_2$, $P_3$, and the drift rate predicted by the model.
One can immediately note several interesting features.
First, the modelling bears out the observation that $P_3$ can indeed change gradually over the course of the mode A drift sequences by a large amount, even by more than a factor of two (from $\sim20$ to $40\,P$ in one case).
Second, $P_2$ appears to be generally smaller for mode B sequences ($\sim12^\circ$) than mode A ($\sim14^\circ$).
Whether or not this is true generally, or whether a single value of $P_2$ could be used to model all drift sequences simultaneously has not been attempted here, but in any case would require a larger sample of drift sequences to test robustly.

\begin{figure}[!th]
    \centering
    \includegraphics[width=\linewidth]{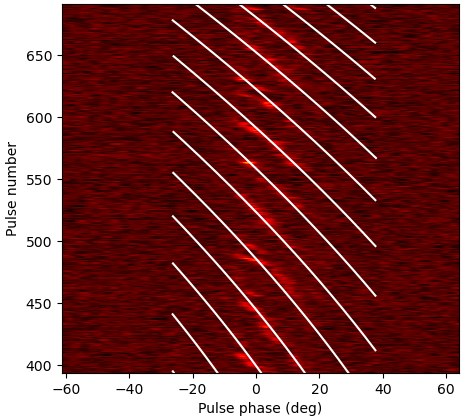}
    \caption{A sequence of mode A2 from the MJD 58434 observation, with fitted model drift band overlaid in white. The ``interruptions'' from the otherwise good fit can be seen most clearly around pulse numbers $\sim450$, $\sim490$, and $\sim620$.}
    \label{fig:A2_example}
\end{figure}

Third, the nature of the disorganization of A2 sequences can now be more closely examined.
Fig. \ref{fig:A2_example} shows such an example A2 sequence to which the above model has been fitted.
Although the model fits well overall, non-random deviations from the fit can be seen in several drift bands.
These deviations seem to consist of an initial rapid increase of the drift rate (i.e.\ the drift bands appear to veer to the left), followed by a short null (only a few pulses long), after which the pulses reappear on the trailing side (i.e.\ the right hand side) of the fitted model before rapidly approaching the fitted model once again.
The temporarily higher drift rate appears visually similar to the drift rate observed during mode B, suggesting that these interruptions are actually brief excursions into mode B.
Curiously, all such deviations (across all observations) appear at about zero phase, which approximately corresponds to the peak of the profile.

\subsubsection{Fluctuation spectra}

\begin{figure}[t]
    \centering
    \includegraphics[width=0.45\textwidth]{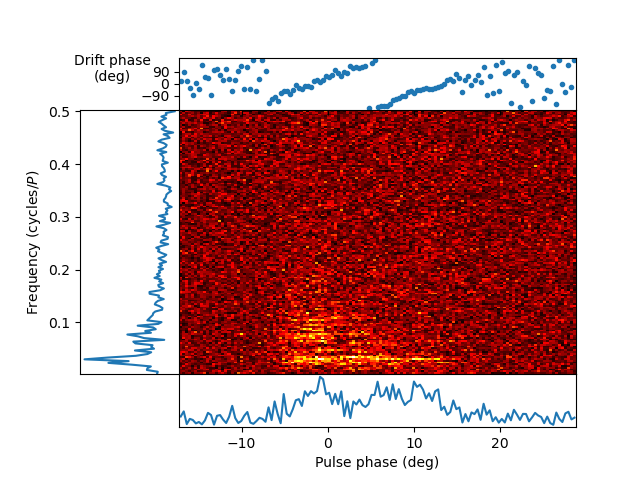} \\
    \includegraphics[width=0.45\textwidth]{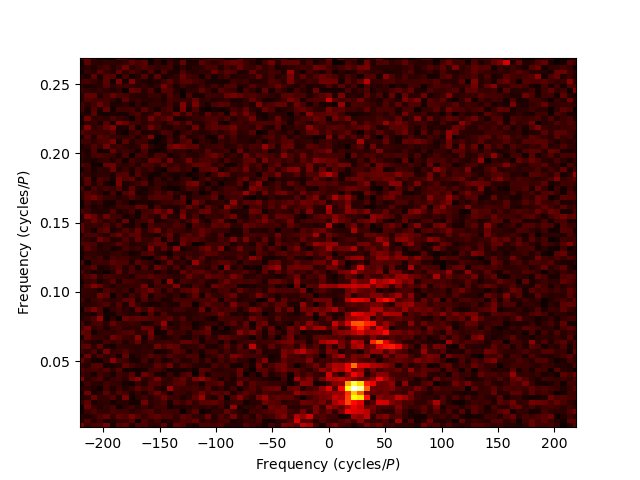}
    \caption{\textit{Top:} The LRFS computed on the section of the pulsestack shown in Fig. \ref{fig:A2_example}. \textit{Bottom:} A zoomed section of the 2DFS of the same pulse stack section. The feature at $\sim0.03\,$cycles/$P$ corresponds to $P_3 \approx 33 \pm 4\,P$, consistent with the average $P_3$ of this sequence determined from the pulse stack directly. The more diffuse power in the range $0.05$-$0.1\,$cycles/$P$ in the leading component (and the corresponding feature visible in the 2DFS) are indicative of the relative disorganization of the drift bands that occurs there.}
    \label{fig:lrfs}
\end{figure}

Fluctuation spectra are most useful when the drift bands are regularly spaced (i.e.\ when $P_3$ is constant).
The fact that this is clearly not the case for \psr{} means that features in the fluctuation spectra will be spread out over a range of spectral bins.
Fig. \ref{fig:lrfs} shows the longitude-resolved fluctuation spectrum (LRFS) and the two-dimensional fluctuation spectrum (2DFS) for the drift sequence shown in Fig. \ref{fig:A2_example}.
Despite the changing $P_3$ and the relative disorganization of the drift rate, the LRFS and 2DFS show surprisingly high-Q features at $P_3 \approx 33 \pm 4\,P$ and $P_2 \approx 16^\circ \pm 2^\circ$, which are consistent with the fitted model values of $P_3$ changing from $\sim50\,P$ to $27\,P$ throughout the sequence, and $P_2 = 13.8^\circ \pm 0.3^\circ$.
The relative lack of power in the bottom panel of the LRFS near phase $\sim2^\circ$ is likely related to the interruptions in the organization of the drift bands that occur at that phase, discussed in the previous section.
Interestingly, the presence of extra power in the LRFS in the leading component between $0.05$-$0.1\,$cycles/$P$ is suggestive of a connection to Mode B, which (with its $P_3 \approx 10\,P$) would be expected to contribute power in the LRFS near this range.

\section{Discussion}
\label{sec:discussion}

PSR \psr{} is a clear example of a pulsar that exhibits sub-pulse drifting; its drifting behavior is qualitatively similar to other pulsars of the same class.
The slope of the drift bands always has the same sign (later sub-pulses arrive at earlier rotation phases than their predecessors), and the drift bands are almost universally connected across the entire pulse window.

These basic facts, when interpreted in the context of the hollow cone/carousel model \citep{Ruderman1975,Rankin1986,Deshpande1999}, suggest that \psr{} is a ``conal single'' pulsar, where the line of sight cuts tangentially through the cone of emission that plays host to a carousel of discrete beams rotating around the star's magnetic axis.
The profiles (see Fig. \ref{fig:profiles}) show a barely-resolved double peak structure, which suggests that the line of sight is just on the poleward side of the emission cone at this frequency.

The most remarkable feature of \psr{}'s drifting behavior is how the drift rate evolves over time, exhibiting both rapid changes (indicating the presence of multiple, distinct drift modes) as well as gradual, intra-mode changes.
The presence of multiple drift modes alone is known to occur in several other pulsars, e.g. B0031$-$07 \citep{Huguenin1970,Joshi2000} and B1944$+$17 \citep{Deich1986,Kloumann2010}, but unlike \psr{}, the drift rates of these pulsars' modes are usually stable enough that the modes can be uniquely characterised by their $P_3$ values.

Because of this inherent stability, analyses of how the drift rate changes within drift modes (or if it changes at all) are relatively rare, but several types of intra-mode evolution of the drift rate have been historically identified.
B0809$+$74, for example, temporarily adopts a slightly lower drift rate immediately following a null, which then relaxes slowly back to the usual, stable rate after a few tens of pulses.
\citet{Lyne1983} demonstrated that this variability is related to the duration of the null sequence preceding it, and that the drift band phase is ``remembered'' across the null.
B0826$-$34, in contrast, has a drift rate that varies slowly but pseudo randomly at all times, making the identification of discrete drift modes impossible \citep[e.g.][]{Esamdin2012}.
More recently, \citet{McSweeney2017} showed that B0031$-$07, which has long been known to exhibit three stable modes, has a measurable slow evolution of the drift rate within individual drift sequences.
The properties of these and a few other, similar pulsars are presented with \psr{} in Table \ref{tab:comparison}.

\begin{deluxetable*}{lcccl}
\tabletypesize{\scriptsize}
\tablewidth{0pt}
\tablecaption{Comparison between properties of several pulsars with slowly changing drift rates \label{tab:comparison}}
\tablehead{
\colhead{Pulsar} & \colhead{Nulling} & \colhead{$P_3$} & \colhead{Relaxation} & \colhead{References} \\[-8pt]
& \colhead{fraction (\%)} & \colhead{($P$)} & time, $\tau_r$ ($P$) &
}
\startdata
\psr{} & 77 & (A) 20-55 & $\gtrsim 100$ & This work \\
& & (B) $\sim10$ & & \\
B0031$-$07 & 45 & (A) $\sim12$ & $\gtrsim 50$ & \citet{Vivekanand1996,Joshi2000}, \\
& & (B) $\sim7$ & & \citep{McSweeney2017,McSweeney2019a} \\
& & (C) $\sim4$ & & \\
B0809$+$74 & 1.4 & $\sim10.8$ & $\sim$10-20 & \citet{Lyne1983,VanLeeuwen2002,VanLeeuwen2003} \\
B0826$-$34 & (weak mode) 70  & Chaotic & - & \citet{Esamdin2005,Esamdin2012,VanLeeuwen2012} \\
B0818$-$13 & 1 & 4.7 & Oscillatory & \citet{Lyne1983} \\
B0943$+$10 & $<1$ & 1.867 & $\sim4000$ & \citet{Deshpande2001,Rankin2003,Bilous2018} \\
B2016$+$28 & $\sim1$ & $\sim3$-$12$ & - & \citet{Taylor1975,Naidu2017}
\enddata
\end{deluxetable*}

Of the pulsars listed in the table, \psr{} is arguably most similar to B0031$-$07, with which it shares (1) the presence of more than one drift mode, (2) the fact that $P_3$ changes slowly throughout individual drift sequences \citep{McSweeney2017}, and (3) a relaxation time (if such a model is even applicable) that is longer than the typical duration of the drift sequences themselves.
This last point is what renders the particular drift band model used in this work unable to measure the relaxation time reliably; in this respect it is not significantly better than the ``quadratic'' model used in \citet{McSweeney2017}, which is equivalent to keeping only the lowest order terms in the Taylor expansion of Eq.~\eqref{eqn:exp_phase}.
Notably, \citet{Joshi2000} report that in B0031$-$07, drift phase is indeed remembered across nulls, as first reported for B0809$+$74 and B0818$-$13 by \citet{Lyne1983}, and a visual inspection of the \psr{} pulse stacks suggest that a similar phase memory is also present here, although a more careful analysis is required before this can be verified.

In general, the connection between nulls and drift rate is well established, but the nature of this relationship varies from pulsar to pulsar (see Table \ref{tab:comparison} for references).
The exponentially decaying drift rate reported for B0809$+$74 has a relaxation time of between 10 and 20 pulses, while B0818$-$13's drift rate varies like a damped oscillator on a similar timescale.
In contrast, fluctuation spectral analyses of B0943$+$10 reveal that its drift rate evolves very slowly after a null, only settling to its stable value after several thousand pulses.
B0826$-$34's drift rate evolves on a faster time scale, but in a pseudo-random manner, not unlike B2016$+$28, whose drift rate can vary on a time scale of a few tens of pulses.

All of these complex drifting behaviors present a natural challenge to the carousel model, whose most basic formulation predicts only a single stable drifting mode, due to the expected stability of the underlying electric and magnetic fields that generate the carousel motion.
Nevertheless, many of the above effects can be successfully explained with the carousel model by invoking aliasing effects that arise due to beating between the stellar rotation rate and the carousel's rotation.
In the presence of aliasing, multiple drift modes can be explained if the number of sparks in the carousel changes, even if the carousel speed itself doesn't change.
This idea has been invoked to explain the multiple drift rates of B1918$+$19 \citep{Rankin2013}, B1819$-$22 \citep{Joshi2017,Janagal2022}, and B0031$-$07 \citep{McSweeney2019a}.
Aliasing can also ``amplify'' the variability of the drift rate if the ratio of stellar rotation and carousel rotation is very close to a ratio of small integers.
That is, a small change in the magnetospheric conditions (e.g. the local electromagnetic configuration) near the stellar surface (in the gap, where the sparks reside) can appear to the observer as a relatively large change in the drift rate.
\citet{VanLeeuwen2012} elucidated the connection between the accelerating potential within the polar gap and the carousel behavior, and found that the variability of B0826$-$34's drift rate can be explained in this way.
In contrast, \citet{VanLeeuwen2003} argued that aliasing could \emph{not} be present during the post-null changing drift rate for B0809$+$74.
Whether or not aliasing is included, models for changing drift rates are often discussed in terms of the local magnetospheric conditions changing over the relevant time scales \citep[e.g.][]{VanLeeuwen2003, Yuen2019}.

One must question whether or not these varied behaviors are all manifestations of the same underlying phenomenon, operating on different timescales.
That is, the general rule that governs all these pulsars might be that the drift rate varies in a systematic way following (and sometimes preceding) a null sequence, and that given ample time before another interruption, all would settle into a stable drift rate.
This is difficult to test for pulsars like \psr{} and B0031$-$07 unless a sufficiently long, uninterrupted drift sequence is observed with a sufficiently short relaxation time.
Among the observations of \psr{}, only one mode A1 drift sequence was found whose fitted relaxation time of $\tau_r \approx 31\,P$ was significantly shorter than the drift sequence itself (285 pulses).
This sequence starts with $P_3 \approx 12\,P$, and after stabilizing, continues on for well over 100 pulses with $P_3 \approx 40\,P$.

Modelling the drift sequences in this way opens up new avenues for exploring and quantifying the relationships between the drift rate, the drift decay rate, the duration of the preceding drift (or null) sequence, the possible phase connection between consecutive drift modes (or across nulls), and others.
For example, in many instances the drift bands appear to connect smoothly across mode switches between modes A and B, which, if true, has implications for the mechanisms by which mode transitions occur in the context of the carousel model.
Related to this is the possibility that the drift rate ``interruptions'' seen in mode A2 sequences are actually miniature excursions into mode B.
If so, why these excursions occur preferentially within a particular phase range also has implications for the carousel model.
The verification and exploration of such claims require much larger data sets than are currently available, and will be explored in a follow-up paper with an enlarged data set comprising more archival MWA observations as well as recently taken uGMRT observations of this pulsar.

\section{Conclusions and summary}
\label{sec:conclusions}

We have reported the independent discovery and first MWA observations of PSR \psr{} in the SMART pulsar survey.
It is bright enough to see in single pulses at MWA frequencies, but contains long null sequences which in some cases can apparently exceed 20 minutes.
Such pulsars are inherently difficult to detect because of their intermittent nature, but are expected to be discovered in greater numbers in surveys like SMART and LOTAAS, which employ long $\ge 1\,$hr dwell times.

Its slow period and intermittent nature mean that long duration observations are required to collect a sufficiently large number of complete burst sequences to undertake robust statistical analysis on their properties.
Follow-up observations are currently under way at the uGMRT and at Murriyang, which will allow us to explore the drifting behavior of \psr{} in much greater depth, as well as obtain a timing solution, refine the nulling fraction estimate, and perform a polarimetric analysis of its single pulses.

However, even with the relatively few observations presently at our disposal, \psr{} clearly exhibits many interesting single pulse behaviors.
Among these, the most prominent is the manner in which the drift rate changes slowly throughout drift sequences (a property observed in only a small number of other sub-pulse drifting pulsars), as well as the interplay between the two drift mode classes identified here (modes A and B) and the intervening null sequences.
These properties make the interpretation of this pulsar's drifting behavior particularly interesting in the context of the carousel model.
In-depth investigations of this kind will thus play an important role in the quest to uncover the intricacies of pulsar emission physics, an outstanding problem in pulsar astronomy.

\begin{acknowledgements}
This scientific work makes use of the Murchison Radio-astronomy Observatory, operated by CSIRO. We acknowledge the Wajarri Yamatji people as the traditional owners of the Observatory site. Support for the operation of the MWA is provided by the Australian Government (NCRIS), under a contract to Curtin University administered by Astronomy Australia Limited.
We acknowledge the Pawsey Supercomputing Centre which is supported by the Western Australian and Australian Governments.
This work was supported by resources awarded under Astronomy Australia Ltd's ASTAC merit allocation scheme on the OzSTAR national facility at the Swinburne University of Technology. The OzSTAR program receives funding in part from the Astronomy National Collaborative Research Infrastructure Strategy (NCRIS) allocation provided by the Australian Government.
The GMRT is run by the National Centre for Radio Astrophysics
of the Tata Institute of Fundamental Research, India.
Software support resources awarded under the Astronomy Data and Computing Services (ADACS) Merit Allocation Program.
ADACS is funded from the Astronomy National Collaborative Research Infrastructure Strategy (NCRIS) allocation provided by the Australian Government and managed by Astronomy Australia Limited (AAL).
DK was supported by NSF grant AST-1816492.
We would also like to thank Alex McEwen for making the original discovery of this pulsar in the GBNCC survey.
RMS acknowledges support through Australian Research Council Future Fellowship FT190100155.
We thank the anonymous referee for the many detailed suggestions which improved this paper.
\end{acknowledgements}

\bibliography{main}{}

\begin{thebibliography}{}
\expandafter\ifx\csname natexlab\endcsname\relax\def\natexlab#1{#1}\fi
\providecommand{\url}[1]{\href{#1}{#1}}
\providecommand{\dodoi}[1]{doi:~\href{http://doi.org/#1}{\nolinkurl{#1}}}
\providecommand{\doeprint}[1]{\href{http://ascl.net/#1}{\nolinkurl{http://ascl.net/#1}}}
\providecommand{\doarXiv}[1]{\href{https://arxiv.org/abs/#1}{\nolinkurl{https://arxiv.org/abs/#1}}}

\bibitem[{Antoniadis {et~al.}(2013)Antoniadis, Freire, Wex, M., S., H.,
  Michael, Cees, S., Thomas, T., M., I., Norbert, R., A., T., M., H., Joeri,
  W., \& G.}]{Antoniadis2013}
Antoniadis, J., Freire, P. C.~C., Wex, N., {et~al.} 2013, Science, 340,
  1233232, \dodoi{10.1126/science.1233232}

\bibitem[{Basu {et~al.}(2020)Basu, Mitra, \& Melikidze}]{Basu2020}
Basu, R., Mitra, D., \& Melikidze, G.~I. 2020, \mnras, 496, 465,
  \dodoi{10.1093/mnras/staa1574}

\bibitem[{Basu {et~al.}(2019)Basu, Mitra, Melikidze, \& Skrzypczak}]{Basu2019a}
Basu, R., Mitra, D., Melikidze, G.~I., \& Skrzypczak, A. 2019, \mnras, 482,
  3757, \dodoi{10.1093/mnras/sty2846}

\bibitem[{Bhat {et~al.}(2018)Bhat, Tremblay, Kirsten, Meyers, Sokolowski, van
  Straten, McSweeney, Ord, Shannon, Beardsley, Crosse, Emrich, Franzen,
  Horsley, Johnston-Hollitt, Kaplan, Kenney, Morales, Pallot, Steele, Tingay,
  Trott, Walker, Wayth, Williams, \& Wu}]{Bhat2018}
Bhat, N. D.~R., Tremblay, S.~E., Kirsten, F., {et~al.} 2018, \apjs, 238,
  \dodoi{10.3847/1538-4365/aad37c}

\bibitem[{Bilous(2018)}]{Bilous2018}
Bilous, A. 2018, \aap, 616, A119, \dodoi{10.1051/0004-6361/201732106}

\bibitem[{Deich {et~al.}(1986)Deich, Cordes, Hankins, \& Rankin}]{Deich1986}
Deich, W., Cordes, J., Hankins, T., \& Rankin, J. 1986, \apj, 300, 540,
  \dodoi{10.1086/163831}

\bibitem[{Demorest {et~al.}(2010)Demorest, Pennucci, Ransom, Roberts, \&
  Hessels}]{Demorest2010}
Demorest, P.~B., Pennucci, T., Ransom, S.~M., Roberts, M. S.~E., \& Hessels, J.
  W.~T. 2010, \nat, 467, 1081, \dodoi{10.1038/nature09466}

\bibitem[{Deshpande \& Rankin(1999)}]{Deshpande1999}
Deshpande, A.~A., \& Rankin, J.~M. 1999, \apj, 524, 1008,
  \dodoi{10.1086/307862}

\bibitem[{Deshpande \& Rankin(2001)}]{Deshpande2001}
---. 2001, \mnras, 322, 438, \dodoi{10.1046/j.1365-8711.2001.04079.x}

\bibitem[{Drake \& Craft(1968)}]{Drake1968}
Drake, F.~D., \& Craft, H.~D. 1968, \nat, 220, 231, \dodoi{10.1038/220231a0}

\bibitem[{Esamdin {et~al.}(2012)Esamdin, Abdurixit, Manchester, \&
  Niu}]{Esamdin2012}
Esamdin, A., Abdurixit, D., Manchester, R.~N., \& Niu, H.~B. 2012, \apj, 759,
  L3, \dodoi{10.1088/2041-8205/759/1/L3}

\bibitem[{Esamdin {et~al.}(2005)Esamdin, Lyne, Graham-Smith, Kramer,
  Manchester, \& Wu}]{Esamdin2005}
Esamdin, A., Lyne, A.~G., Graham-Smith, F., {et~al.} 2005, \mnras, 356, 59,
  \dodoi{10.1111/j.1365-2966.2004.08444.x}

\bibitem[{Gao {et~al.}(2017)Gao, Wang, Shan, Li, \& Wang}]{Gao2017}
Gao, Z.-F., Wang, N., Shan, H., Li, X.-D., \& Wang, W. 2017, \apj, 849, 19,
  \dodoi{10.3847/1538-4357/aa8f49}

\bibitem[{Gil \& Sendyk(2000)}]{Gil2000}
Gil, J.~A., \& Sendyk, M. 2000, \apj, 541, 351, \dodoi{10.1086/309394}

\bibitem[{Gupta {et~al.}(2017)Gupta, Ajithkumar, Kale, Nayak, Sabhapathy,
  Sureshkumar, Swami, Chengalur, Ghosh, Ishwara-Chandra, Joshi, Kanekar, Lal,
  \& Roy}]{Gupta2017}
Gupta, Y., Ajithkumar, B., Kale, H.~S., {et~al.} 2017, Curr. Sci, 113, 707

\bibitem[{Hewish {et~al.}(1968)Hewish, Bell, Pilkington, Scott, \&
  Collins}]{Hewish1968}
Hewish, A., Bell, S.~J., Pilkington, J. D.~H., Scott, P.~F., \& Collins, R.~A.
  1968, \nat, 217, 709, \dodoi{10.1038/224472b0}

\bibitem[{Hotan {et~al.}(2004)Hotan, {Van Straten}, \& Manchester}]{Hotan2004}
Hotan, a.~W., {Van Straten}, W., \& Manchester, R.~N. 2004, \pasa, 21, 302,
  \dodoi{10.1071/AS04022}

\bibitem[{Huguenin {et~al.}(1970)Huguenin, Taylor, \& Troland}]{Huguenin1970}
Huguenin, G.~R., Taylor, J.~H., \& Troland, T.~H. 1970, \apj, 162, 727,
  \dodoi{10.1086/150704}

\bibitem[{Janagal {et~al.}(2022)Janagal, Chakraborty, Bhat, Bhattacharyya, \&
  McSweeney}]{Janagal2022}
Janagal, P., Chakraborty, M., Bhat, N. D.~R., Bhattacharyya, B., \& McSweeney,
  S.~J. 2022, \mnras, 509, 4573, \dodoi{10.1093/mnras/stab3305}

\bibitem[{Joshi {et~al.}(2017)Joshi, Naidu, Gajjar, \& Wright}]{Joshi2017}
Joshi, B.~C., Naidu, A., Gajjar, V., \& Wright, G. A.~E. 2017, Proc Int Astron,
  13, 348, \dodoi{10.1017/S1743921317008390}

\bibitem[{Joshi \& Vivekanand(2000)}]{Joshi2000}
Joshi, B.~C., \& Vivekanand, M. 2000, \mnras, 316, 716.
\newblock
  \url{http://adsabs.harvard.edu/cgi-bin/nph-bib_query?bibcode=2000MNRAS.316..716J&db_key=AST}

\bibitem[{Kaur {et~al.}(2019)Kaur, Bhat, Tremblay, Shannon, McSweeney, Ord,
  Beardsley, Crosse, Emrich, Franzen, Horsley, Johnston-Hollitt, Kaplan,
  Kenney, Morales, Pallot, Steele, Tingay, Trott, Walker, Wayth, Williams, \&
  Wu}]{Kaur2019}
Kaur, D., Bhat, N. D.~R., Tremblay, S.~E., {et~al.} 2019, \apj, 882, 133,
  \dodoi{10.3847/1538-4357/ab338f}

\bibitem[{Keith {et~al.}(2010)Keith, Jameson, van Straten, Bailes, Johnston,
  Kramer, Possenti, Bates, Bhat, Burgay, Burke-Spolaor, D'Amico, Levin,
  McMahon, Milia, \& Stappers}]{Keith2010}
Keith, M.~J., Jameson, A., van Straten, W., {et~al.} 2010, \mnras, 409, 619,
  \dodoi{10.1111/j.1365-2966.2010.17325.x}

\bibitem[{Kloumann \& Rankin(2010)}]{Kloumann2010}
Kloumann, I.~M., \& Rankin, J.~M. 2010, \mnras, 408, 40,
  \dodoi{10.1111/j.1365-2966.2010.17114.x}

\bibitem[{Kramer {et~al.}(2006)Kramer, Stairs, Manchester, McLaughlin, Lyne,
  Ferdman, Burgay, Lorimer, Possenti, D'Amico, Sarkissian, Hobbs, Reynolds,
  Freire, \& Camilo}]{Kramer2006}
Kramer, M., Stairs, I.~H., Manchester, R.~N., {et~al.} 2006, Science, 314, 97,
  \dodoi{10.1126/science.1132305}

\bibitem[{Lyne \& Ashworth(1983)}]{Lyne1983}
Lyne, A.~G., \& Ashworth, M. 1983, \mnras, 204, 519,
  \dodoi{10.1093/mnras/204.2.519}

\bibitem[{Manchester {et~al.}(1996)Manchester, Lyne, D'Amico, Bailes, Johnston,
  Lorimer, Harrison, Nicastro, \& Bell}]{Manchester1996}
Manchester, R., Lyne, A., D'Amico, N., {et~al.} 1996, \mnras, 279, 1235

\bibitem[{Manchester {et~al.}(2013)Manchester, Hobbs, Bailes, Coles, van
  Straten, Keith, Shannon, Bhat, Brown, Burke-Spolaor, Champion, Chaudhary,
  Edwards, Hampson, Hotan, Jameson, Jenet, Kesteven, Khoo, Kocz, Maciesiak,
  Oslowski, Ravi, Reynolds, Sarkissian, Verbiest, Wen, Wilson, Yardley, Yan, \&
  You}]{Manchester2013}
Manchester, R.~N., Hobbs, G., Bailes, M., {et~al.} 2013, \pasa, 30, e017,
  \dodoi{10.1017/pasa.2012.017}

\bibitem[{McSweeney {et~al.}(2017)McSweeney, Bhat, Tremblay, Deshpande, \&
  Ord}]{McSweeney2017}
McSweeney, S.~J., Bhat, N. D.~R., Tremblay, S.~E., Deshpande, A.~A., \& Ord,
  S.~M. 2017, \apj, 836, 224, \dodoi{10.3847/1538-4357/aa5c35}

\bibitem[{McSweeney {et~al.}(2019)McSweeney, Bhat, Wright, Tremblay, \&
  Kudale}]{McSweeney2019a}
McSweeney, S.~J., Bhat, N. D.~R., Wright, G., Tremblay, S.~E., \& Kudale, S.
  2019, \apj, 883, 28, \dodoi{10.3847/1538-4357/ab3a97}

\bibitem[{McSweeney {et~al.}(2020)McSweeney, Ord, Kaur, Bhat, Meyers, Tremblay,
  Jones, Crosse, \& Smith}]{McSweeney2020}
McSweeney, S.~J., Ord, S.~M., Kaur, D., {et~al.} 2020, \pasa, 37, e034,
  \dodoi{10.1017/pasa.2020.24}

\bibitem[{Melrose {et~al.}(2021)Melrose, Rafat, \& Mastrano}]{Melrose2021}
Melrose, D.~B., Rafat, M.~Z., \& Mastrano, A. 2021, \mnras, 500, 4530,
  \dodoi{10.1093/mnras/staa3324}

\bibitem[{Meyers {et~al.}(2017)Meyers, Tremblay, Bhat, Shannon, Kirsten,
  Sokolowski, Tingay, Oronsaye, \& Ord}]{Meyers2017}
Meyers, B.~W., Tremblay, S.~E., Bhat, N. D.~R., {et~al.} 2017, \apj, 851, 20,
  \dodoi{10.3847/1538-4357/aa8bba}

\bibitem[{Miao {et~al.}(2021)Miao, Xu, Shao, Liu, \& Ma}]{Miao2021}
Miao, X., Xu, H., Shao, L., Liu, C., \& Ma, B.-Q. 2021, \apj, 921, 114,
  \dodoi{10.3847/1538-4357/ac1d48}

\bibitem[{Mitchell {et~al.}(2008)Mitchell, Greenhill, Wayth, Sault, Lonsdale,
  Cappallo, Morales, \& Ord}]{Mitchell2008}
Mitchell, D.~A., Greenhill, L.~J., Wayth, R.~B., {et~al.} 2008, IEEE J. Sel, 2,
  707, \dodoi{10.1109/JSTSP.2008.2005327}

\bibitem[{Naidu {et~al.}(2017)Naidu, Joshi, Manoharan, \&
  KrishnaKumar}]{Naidu2017}
Naidu, A., Joshi, B.~C., Manoharan, P.~K., \& KrishnaKumar, M.~A. 2017, \aap,
  604, A45, \dodoi{10.1051/0004-6361/201629937}

\bibitem[{Ord {et~al.}(2019)Ord, Tremblay, McSweeney, Bhat, Sobey, Mitchell,
  Hancock, \& Kirsten}]{Ord2019}
Ord, S.~M., Tremblay, S.~E., McSweeney, S.~J., {et~al.} 2019, \pasa, 36, e030,
  \dodoi{10.1017/pasa.2019.17}

\bibitem[{Pang {et~al.}(2021)Pang, Tews, Coughlin, Bulla, {Van Den Broeck}, \&
  Dietrich}]{Pang2021}
Pang, P. T.~H., Tews, I., Coughlin, M.~W., {et~al.} 2021, \apj, 922, 14,
  \dodoi{10.3847/1538-4357/ac19ab}

\bibitem[{Qiao {et~al.}(2004)Qiao, Lee, Zhang, Xu, \& Wang}]{Qiao2004}
Qiao, G.~J., Lee, K.~J., Zhang, B., Xu, R.~X., \& Wang, H.~G. 2004, \apj, 616,
  L127, \dodoi{10.1086/426862}

\bibitem[{Rankin(1986)}]{Rankin1986}
Rankin, J.~M. 1986, \apj, 301, 901, \dodoi{10.1086/163955}

\bibitem[{Rankin {et~al.}(2003)Rankin, Suleymanova, \& Deshpande}]{Rankin2003}
Rankin, J.~M., Suleymanova, S.~a., \& Deshpande, A.~a. 2003, \mnras, 340, 1076,
  \dodoi{10.1046/j.1365-8711.2003.06391.x}

\bibitem[{Rankin {et~al.}(2013)Rankin, Wright, \& Brown}]{Rankin2013}
Rankin, J.~M., Wright, G. A.~E., \& Brown, A.~M. 2013, \mnras, 433, 445,
  \dodoi{10.1093/mnras/stt739}

\bibitem[{Ruderman \& Sutherland(1975)}]{Ruderman1975}
Ruderman, M.~A., \& Sutherland, P.~G. 1975, \apj, 196, 51,
  \dodoi{10.1086/153393}

\bibitem[{Sanidas {et~al.}(2019)Sanidas, Cooper, Bassa, Hessels, Kondratiev,
  Michilli, Stappers, Tan, van Leeuwen, Cerrigone, Fallows, Iacobelli,
  Orr{\'{u}}, Pizzo, Shulevski, Toribio, ter Veen, Zucca, Bondonneau,
  Grie{\ss}meier, Karastergiou, Kramer, \& Sobey}]{Sanidas2019}
Sanidas, S., Cooper, S., Bassa, C., {et~al.} 2019, \aap, 626, A104,
  \dodoi{10.1051/0004-6361/201935609}

\bibitem[{Stovall {et~al.}(2014)Stovall, Lynch, Ransom, Archibald, Banaszak,
  Biwer, Boyles, Dartez, Day, Ford, Flanigan, Garcia, Hessels, Hinojosa, Jenet,
  Kaplan, Karako-Argaman, Kaspi, Kondratiev, Leake, Lorimer, Lunsford,
  Martinez, Mata, McLaughlin, Roberts, Rohr, Siemens, Stairs, van Leeuwen,
  Walker, \& Wells}]{Stovall2014}
Stovall, K., Lynch, R.~S., Ransom, S.~M., {et~al.} 2014, \apj, 791, 67,
  \dodoi{10.1088/0004-637X/791/1/67}

\bibitem[{Swainston {et~al.}(2022)Swainston, Bhat, Morrison, McSweeney, Ord,
  Tremblay, \& Sokolowski}]{Swainston2022}
Swainston, N.~A., Bhat, N. D.~R., Morrison, I.~S., {et~al.} 2022, \pasa, 39,
  e020, \dodoi{DOI: 10.1017/pasa.2022.14}

\bibitem[{Swainston {et~al.}(2021)Swainston, Bhat, Sokolowski, McSweeney,
  Kudale, Dai, Smith, Morrison, Shannon, van Straten, Xue, Ord, Tremblay,
  Meyers, Williams, Sleap, Johnston-Hollitt, Kaplan, Tingay, \&
  Wayth}]{Swainston2021}
Swainston, N.~A., Bhat, N. D.~R., Sokolowski, M., {et~al.} 2021, \apjl, 911,
  L26, \dodoi{10.3847/2041-8213/abec7b}

\bibitem[{Taylor {et~al.}(1971)Taylor, Huguenin, Hirsch, \&
  Manchester}]{Taylor1971}
Taylor, J.~H., Huguenin, G.~R., Hirsch, R.~M., \& Manchester, R.~N. 1971,
  \apjl, 9, 205

\bibitem[{Taylor {et~al.}(1975)Taylor, Manchester, \& Huguenin}]{Taylor1975}
Taylor, J.~H., Manchester, R.~N., \& Huguenin, G.~R. 1975, \apj, 195, 513,
  \dodoi{10.1086/153351}

\bibitem[{Tingay {et~al.}(2013)Tingay, Goeke, Bowman, Emrich, Ord, Mitchell,
  Morales, Booler, Crosse, Wayth, Lonsdale, Tremblay, Pallot, Colegate,
  Wicenec, Kudryavtseva, Arcus, Barnes, Bernardi, Briggs, Burns, Bunton,
  Cappallo, Corey, Deshpande, DeSouza, Gaensler, Greenhill, Hall, Hazelton,
  Herne, Hewitt, Johnston-Hollitt, Kaplan, Kasper, Kincaid, Koenig,
  Kratzenberg, Lynch, McKinley, McWhirter, Morgan, Oberoi, Pathikulangara,
  Prabu, Remillard, Rogers, Roshi, Salah, Sault, Udaya-Shankar, Schlagenhaufer,
  Srivani, Stevens, Subrahmanyan, Waterson, Webster, Whitney, Williams,
  Williams, \& Wyithe}]{Tingay2013}
Tingay, S.~J., Goeke, R., Bowman, J.~D., {et~al.} 2013, \pasa, 30, 7,
  \dodoi{10.1017/pasa.2012.007}

\bibitem[{Tremblay {et~al.}(2015)Tremblay, Ord, Bhat, Tingay, Crosse, Pallot,
  Oronsaye, Bernardi, Bowman, Briggs, Cappallo, Corey, Deshpande, Emrich,
  Goeke, Greenhill, Hazelton, Johnston-Hollitt, Kaplan, Kasper, Kratzenberg,
  Lonsdale, Lynch, McWhirter, Mitchell, Morales, Morgan, Oberoi, Prabu, Rogers,
  Roshi, {Udaya Shankar}, Srivani, Subrahmanyan, Waterson, Wayth, Webster,
  Whitney, Williams, \& Williams}]{Tremblay2015}
Tremblay, S.~E., Ord, S.~M., Bhat, N. D.~R., {et~al.} 2015, \pasa, 32, e005,
  \dodoi{10.1017/pasa.2015.6}

\bibitem[{van Leeuwen {et~al.}(2002)van Leeuwen, Kouwenhoven, Ramachand~ran,
  Rankin, \& Stappers}]{VanLeeuwen2002}
van Leeuwen, A., Kouwenhoven, M., Ramachand~ran, R., Rankin, J., \& Stappers,
  B. 2002, \aap, 387, 169, \dodoi{10.1051/0004-6361:20020254}

\bibitem[{van Leeuwen {et~al.}(2003)van Leeuwen, Stappers, Ramachandran, \&
  Rankin}]{VanLeeuwen2003}
van Leeuwen, A. G.~J., Stappers, B.~W., Ramachandran, R., \& Rankin, J.~M.
  2003, \aap, 399, 223, \dodoi{10.1051/0004-6361}

\bibitem[{van Leeuwen \& Timokhin(2012)}]{VanLeeuwen2012}
van Leeuwen, J., \& Timokhin, A.~N. 2012, \apj, 752, 155,
  \dodoi{10.1088/0004-637x/752/2/155}

\bibitem[{van Straten \& Bailes(2011)}]{VanStraten2011b}
van Straten, W., \& Bailes, M. 2011, \pasa, 28, 1, \dodoi{10.1071/AS10021}

\bibitem[{Vivekanand \& Joshi(1996)}]{Vivekanand1996}
Vivekanand, M., \& Joshi, B.~C. 1996, \apj, 477, 431, \dodoi{10.1086/303690}

\bibitem[{Wang {et~al.}(2020)Wang, Gao, Jia, Wang, \& Li}]{Wang2020}
Wang, H., Gao, Z.-F., Jia, H.-Y., Wang, N., \& Li, X.-D. 2020, {Estimation of
  Electrical Conductivity and Magnetization Parameter of Neutron Star Crusts
  and Applied to the High-Braking-Index Pulsar PSR J1640-4631},
  \dodoi{10.3390/universe6050063}

\bibitem[{Wayth {et~al.}(2018)Wayth, Tingay, Trott, Emrich, Johnston-Hollitt,
  McKinley, Gaensler, Beardsley, Booler, Crosse, Franzen, Horsley, Kaplan,
  Kenney, Morales, Pallot, Sleap, Steele, Walker, Williams, Wu, Cairns,
  Filipovic, Johnston, Murphy, Quinn, Staveley-Smith, Webster, \&
  Wyithe}]{Wayth2018}
Wayth, R.~B., Tingay, S.~J., Trott, C.~M., {et~al.} 2018, \pasa, 35, 33,
  \dodoi{10.1017/pasa.2018.37}

\bibitem[{Wen {et~al.}(2016)Wen, Wang, Yuan, Yan, Manchester, Yuen, \&
  Gajjar}]{Wen2016}
Wen, Z.~G., Wang, N., Yuan, J.~P., {et~al.} 2016, \aap, 592, A127,
  \dodoi{10.1051/0004-6361/201628214}

\bibitem[{Wen {et~al.}(2020)Wen, Chen, Hao, Yan, Wang, Li, Yuan, Lee, Wang,
  Yuen, Xu, Li, \& Huang}]{Wen2020}
Wen, Z.~G., Chen, J.~L., Hao, L.~F., {et~al.} 2020, \apj, 900, 168,
  \dodoi{10.3847/1538-4357/abaab7}

\bibitem[{Wright \& Weltevrede(2017)}]{Wright2017}
Wright, G., \& Weltevrede, P. 2017, \mnras, 464, 2597,
  \dodoi{10.1093/mnras/stw2498}

\bibitem[{Yuen(2019)}]{Yuen2019}
Yuen, R. 2019, \mnras, 486, 2011, \dodoi{10.1093/mnras/stz951}

\end{thebibliography}
\bibliographystyle{aasjournal}



\end{document}